# ASE suppression in $Er^{3+}$ doped dual-core triangular lattice Photonic Crystal Fibers (PCFs) for communication wavelength

Partha Sona Maji* and Partha Roy Chaudhuri

Department of Physics & Meteorology, Indian Institute of Technology Kharagpur-721 302, INDIA

*Tel: +91-3222-283842 Fax: +91-3222-255303,*

**Corresponding author:* parthamaji@phy.iitkgp.ernet.in

**Abstract:** In this article, silica based triangular lattice PCF has been investigated towards both narrowband and broadband dispersion compensation for application in the communication wavelength. A dual core structure is obtained by introducing two different air-hole diameters in the cladding of the PCF. Dependence of individual structural parameters towards high negative dispersion (both narrowband and broadband) has been investigated in details with multipole mode based solver. The numerical investigation exhibits narrowband of very large negative dispersion of −37,300 *ps/nm/km* around the wavelength of 1550 nm. Present investigation also reports broadband dispersion values varying from −800 *ps/nm/km* to −2600 *ps/nm/km* over a 200 nm wavelength (1400 nm to 1600 nm) range, and kappa values near 300 nm, which matches well with standard single mode fiber. Using the principle of power transfer from the inner core to the outer core after the coupling wavelength, we have investigated possible design of ASE suppressed amplifier in which wavelengths after the coupling wavelength cannot be amplified as most of the power tunnel to the outer core, where doped ion does not exist.

## 1. INTRODUCTION

Dispersion which is caused by inline standard Single Mode Fiber (SMF) needs to be compensated as it restricts the bandwidth of the communication channel and obstructs appropriate reproduction of the signal at the receiver end. The typical magnitude of the dispersion parameter varies in the range of $D$=10–20 *ps/km/nm*, with a dispersion slope around 0.05 $ps/nm^2/km$. A variety of different dispersion compensating fibers (DCFs) has been studied to achieve dispersion compensation around the communication wavelength band [1-10]. All solid dual core fibers [1-5], Bragg fibers [6-7], and higher order mode [8-10] based DCFs designs are studied to achieve the target of dispersion compensation.

Though some of these designs can provide higher negative dispersion values [6-7], however the technology involved remain far more complex. To minimize the propagation loss and to reduce the cost of the dispersion compensating modules, the length of the DCFs should be as short as possible with very large magnitude of negative dispersion. In order to achieve an efficient compensation of the dispersion of all the frequencies of Dense Wavelength Division Multiplexing (DWDM) system, the negative dispersion of DCFs should cover a wide spectrum. Thus, an efficient compensation should deal with both dispersion and dispersion slope and thereby matching the kappa values of the DCFs and the inlaying fibers. Therefore, while designing a DCF; one needs to be particularly careful with some of the important parameters such as dispersion, dispersion slope, both the kappa values, bandwidth of the spectrum, and mode property.

The basic principle of DCFs of mode coupling between the two cores can be implemented in Photonic Crystal Fibers (PCFs) by proper adjustment of available different parameters. Conventional PCFs [11-12] have cladding structures formed by air-holes with the *same* diameter arranged in a *regular* triangular or square lattice. By varying the air-hole diameter ($d$) and hole-to-hole spacing ($\Lambda$) of a PCF, the modal properties, in particular, the dispersion properties can easily be engineered. However, a high magnitude of negative dispersion in the target wavelength cannot be achieved with conventional PCFs with all the air-holes having same diameters. Though several works based on PCFs for dispersion compensation have been reported [13-17], until now, as far we are concerned, no works considered both narrow band and broadband dispersion, which we did. Also in the present work, we achieved larger negative dispersion values than the one published so far with the regular dual-core PCFs.

Using the principle of DCFs, we worked out a dual-concentric-core PCF based on pure silica. A relation between dispersion and the structure parameters is obtained. Adjusting the structure parameters of the fiber, we have designed a narrowband DCF with large negative dispersion and a broadband DCF with proper kappa values. To the best of our knowledge, the value we obtained is the largest negative value with a regular triangular lattice PCF. Our numerical investigation establishes that while drawing the fiber, even if there may be some changes in structural parameters, the designed DCFs still demonstrate admirable broadband dispersion-compensating properties.

Subsequently, based on the principle of power transfer from inner core to the outer core at the coupling wavelength, we have designed a possible ASE suppressed amplifier, where wavelengths after the coupling value will not be amplified as most of the power tunnels to the outer-core where amplifying Erbium ions do not exist.

## 2. Geometry of the structure and analysis method:

Schematic diagram of the dual-core concentric PCF has been shown in Fig. 1. The whole fiber is based on pure silica. The central solid region is the inner core. The third ring air-holes are relatively small, locally raising the effective refractive index and forming an annular outer core. Like regular triangular-lattice PCF, the hole to hole distance is "Λ" and the diameter of the larger air-holes is $d_1$; while the diameter of the smaller air-holes (the third ring holes) is $d_2$.

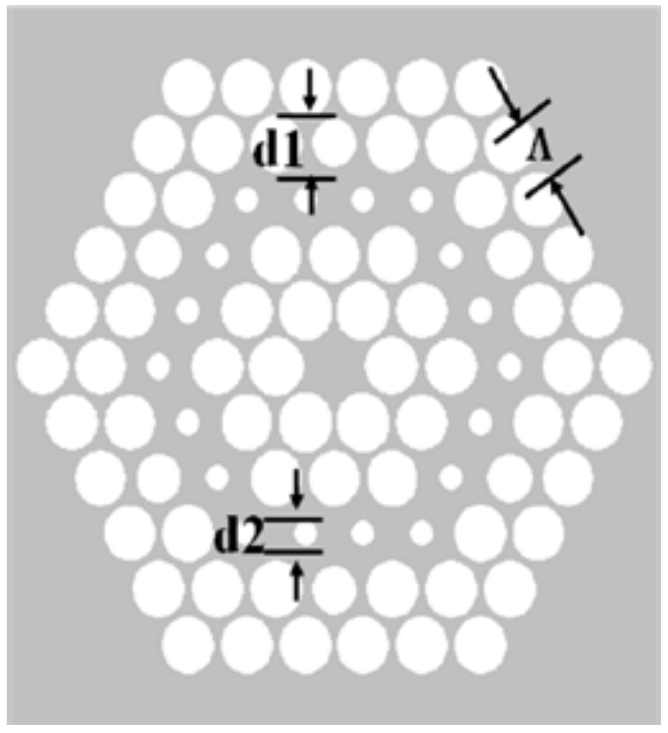

Fig. 1: Cross section of the studied dual core concentric photonic crystal fiber.

We solve the guided modes of the present fiber by the CUDOS MOF Utilities [18] that simulate PCFs using the multipole method [19-20]. The dispersion parameter $D$ is calculated with Eqn. 1 as follows.

$$D = -\frac{\lambda}{c}\frac{d^2 \,\mathrm{Re}[n_{eff}]}{d\lambda^2} \qquad (1)$$

with $Re(n_{eff})$ is the real part of the effective indices obtained from simulations and c is the speed of light in vacuum.

The confinement loss for the structures has been calculated using Eqn. (2).

$$L = \frac{2\pi}{\lambda}\frac{20}{\ln(10)}10^6 \,\mathrm{Im}(n_{eff}) \; dB/m \quad (2)$$

where $\mathrm{Im}(n_{eff})$ is the imaginary part of the effective indices (obtained from the simulations) and λ in micrometer.

We define a new parameter called kappa as follows

$$kappa = \frac{D}{D_s} \qquad (3)$$

with $D_s$ as the dispersion slope. *Kappa* denotes the capability to compensate the dispersion and its slope at the same time.

## 3: Dispersion Analysis of Dual-Concentric-Core

We started our analysis with Λ=1.20 μm, $d_1$= 0.96 μm, and $d_2$=0.48 μm. The modal effective index for the structure for both the fundamental mode and outer core mode has been shown in Fig. 2 and the corresponding $\mathrm{Im}(n_{eff})$ (the loss equivalent as presented by Eqn. 2.) of the guided modes are presented in Fig. 3. In Fig. 3 the two lines corresponding to inner core and outer core intersect each other at the coupling wavelength $\lambda_c$. Figure 4 shows the field distribution of the fundamental mode for the three wavelength region, first one prior to coupling wavelength (1.50 μm), second one during coupling (λ =1.54 μm) and third one after coupling wavelength (λ=1.58 μm). It can be observed from Fig. 4(a) that the inner fundamental mode is well confined in the core with the entire power remaining in the central region. Due to the anti-crossing of the two individual (inner core and outer core) modes, around the coupling wavelength ($\lambda_c$), the $n_{eff}$'s of the two modes come closer to each other and the coupling between the modes can take place as can be seen from Fig. 3 and a rapid change in the slope of the effective index curve can be observed as shown in Fig. 2. Figure 3 shows that at $\lambda_{c.}$ the loss of the two curves cross each other and there is a sharp change in losses in the two curves and after $\lambda_c$ the loss property reverses between the inner mode and outer mode. Figure 4(b) shows that around this coupling wavelength, the field starts to spread out from inner core to the outer core. For $\lambda > \lambda_{c,}$ most of the powers spreads from inner to the outer core and is well confined there as shown in Fig. 4(c). The distinction in the effective indices slope results in a large negative dispersion value around the center of the communication wavelength as shown in Fig. 5. It should be mentioned here that the simulation establishes both the fundamental mode and the outer core mode existing in all the wavelengths for dual-core PCF structures. The fundamental inner core mode will exhibit large negative dispersion as shown in Fig. 5. However at the same time the outer core mode will exhibit a symmetric but reverse replica with equal amount of positive dispersion. To achieve very high negative dispersion we need to excite selectively the inner core fundamental one only. So it is important to excite the correct mode we want. If we can launch a fiber with a mode field very close to the inner core mode, we can let the inner mode carry most of the transmitted power to achieve our goal. This is similar to exciting the circular fundamental mode of a fiber. As a consequence, the outer core mode will not carry any significant power; therefore its effect can be neglected. With this understanding we only discuss the property of the inner core fundamental mode for the article.

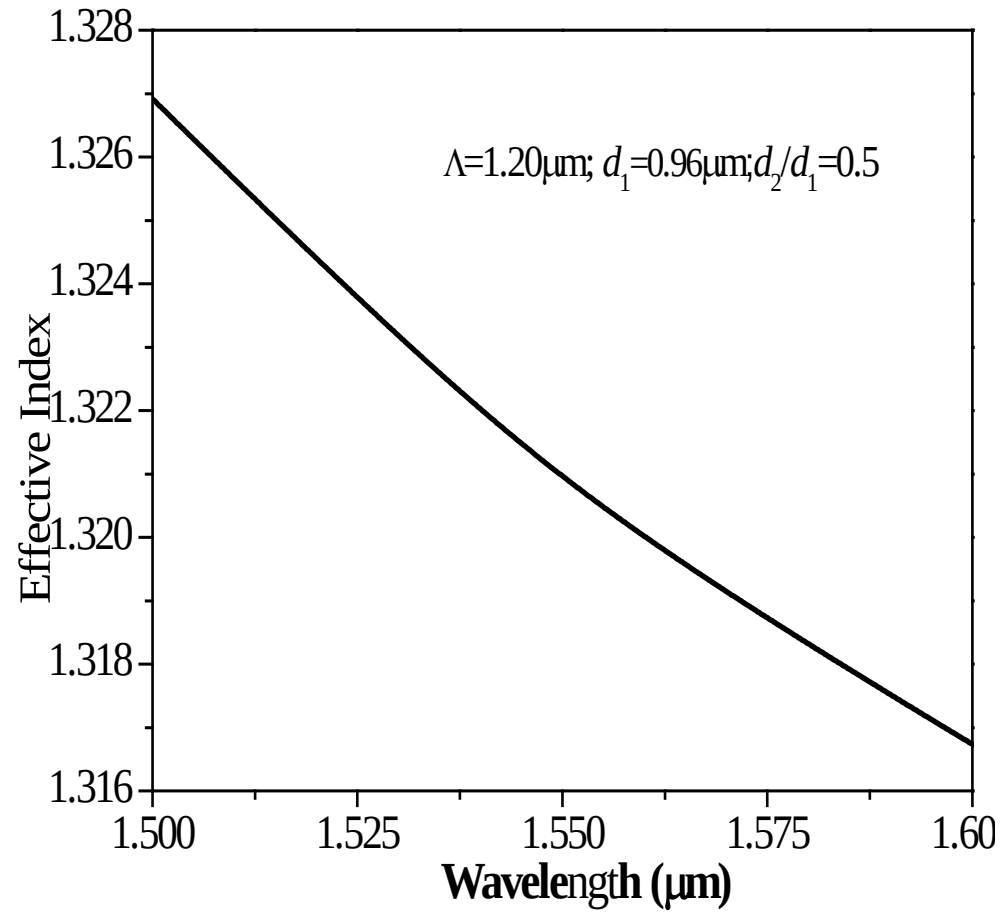

Fig. 2: Effective indices for the fundamental mode with Λ=1.2μm, $d_1$= 0.96μm, and $d_2$=0.48μm.

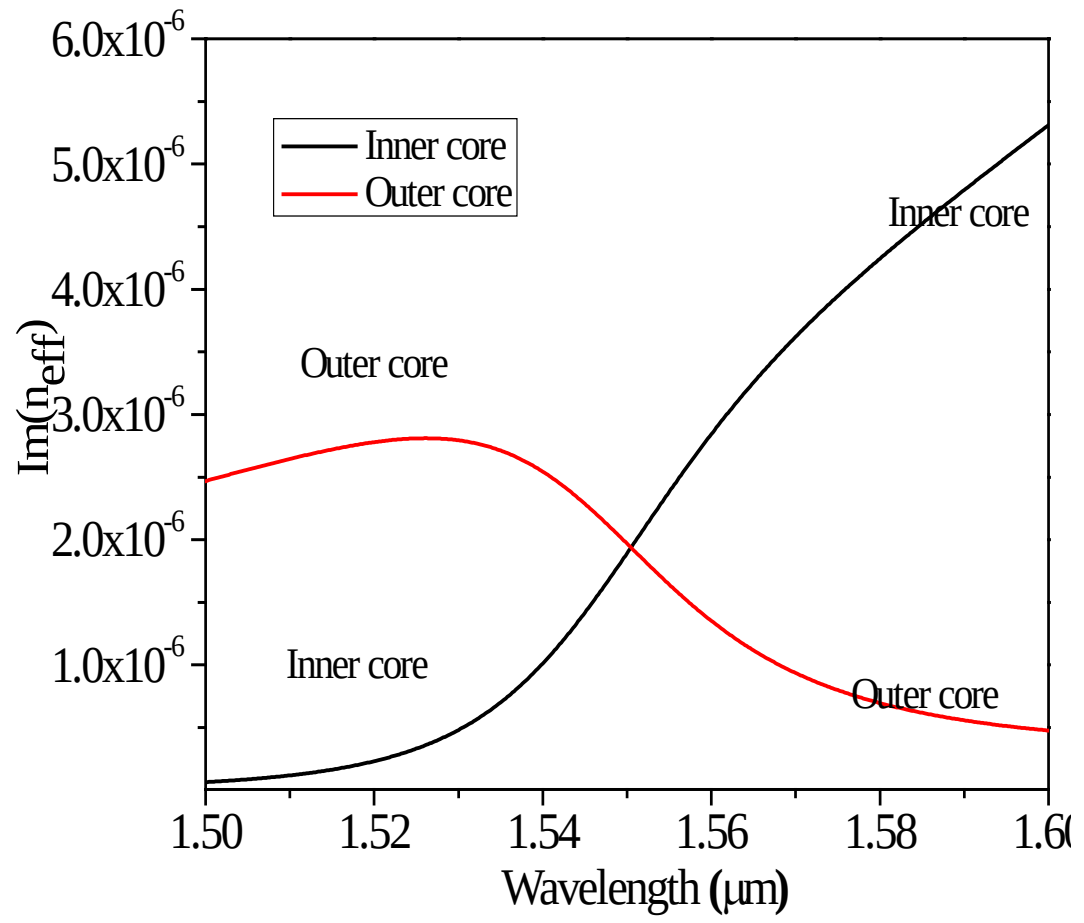

Fig. 3: The Im(neff) (loss equivalent) variation of the two modes (in linear scale) corresponding to Fig. 2.

In the following section, the relationship between dispersion and structural parameters is presented with Figs. 6-8. The effects of variation of bigger air-holes ($d_1$) are represented in Fig. 6 (a). With Λ=1.20 µm, $d_2/d_1$=0.5, the absolute values of the biggest negative dispersion increase as $d_1$ increases (air filling rate increases), the corresponding $\lambda_c$ is red-shifted, and the absolute values of the dispersion slope increase. The typical behavior can be analyzed based on Fig. 6(b) which indicates that for higher values of bigger air-hole diameter the effective indices decreases leading to phase matching wavelength shifting towards higher wavelength. Besides, smaller refractive indices increases the depth of the refractive index profile and thereby suppress the occurrence of the optical coupling between the two cores, resulting in more incisive tips on the index curves and narrower FWHM. Thus, *increasing d has the effect of red-shifting the coupling wavelength with higher negative dispersion, narrower FWHM and vice versa*. The dispersion curves of PCFs for different $d_2$ are shown in Fig. 7(a). With Λ=1.20 µm, $d_1$ =0.96 µm, the absolute values of the biggest negative dispersion decrease as $d_2$ increases (air filling rate of the outer core increases), the corresponding wavelength is red-shifted, and the absolute values of the dispersion slope decrease. This typical nature can be explained again on the basis of effective index curve of Fig. 7(b). For lesser values of small air-hole diameter, the effective index difference between the two cores of the PCFs decreases resulting in phase matching wavelength shifting towards smaller wavelength with perceptive tips. Also the visible sharp changes of effective indices graphs at smaller wavelength results in sharper peak in the dispersion profile and smaller FWHM. Hence, *with the decrease of small air-hole diameter the negative dispersion gets blue-shifted with higher negative dispersion along with a narrower value of FWHM and vice versa*. The effect of changing Λ is presented in Fig. 8. It can be observed with $d_1$/Λ=0.8 and $d_2/d_1$=0.5 (air-filling rate is fixed) that the absolute values of the biggest negative dispersion decrease appreciably as Λ increases, the corresponding wavelength is red-shifted and the absolute values of the dispersion slope decrease appreciably. This can be attributed to the case that increasing the values of Λ increases the distances between the two cores resulting in appearances of rapid slope change of the coupling towards higher wavelength. So, *increasing Λ has the effect of red-shifting the coupling wavelength with lower dispersion and vice versa*. From the above study, we can see, as absolute values of the biggest negative dispersion increase, absolute values of the dispersion slope increase and the width of the dispersion curve loop decrease. In contrast, as absolute values of the biggest negative dispersion decrease, absolute values of the dispersion slope decrease, and the width of the dispersion curve loop increases. Despite the changes in the structure of the fibers, the product of the absolute values of dispersion and the width of the dispersion curve loop are almost invariable. So in practice, we can design narrowband DCPCF with large negative dispersion and broadband DCPCF with proper kappa values by adjusting the structure parameters of PCF.

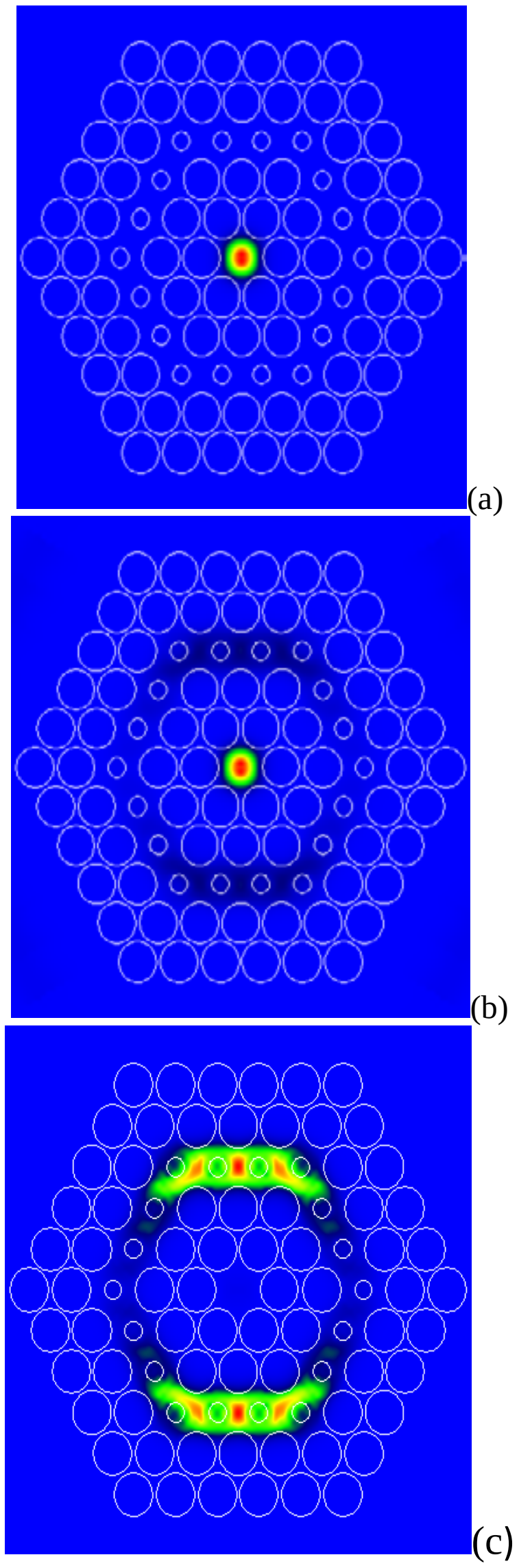

Fig. 4: Field distribution of the fundamental mode (a) long before coupling (b) during coupling (c) after coupling respectively.

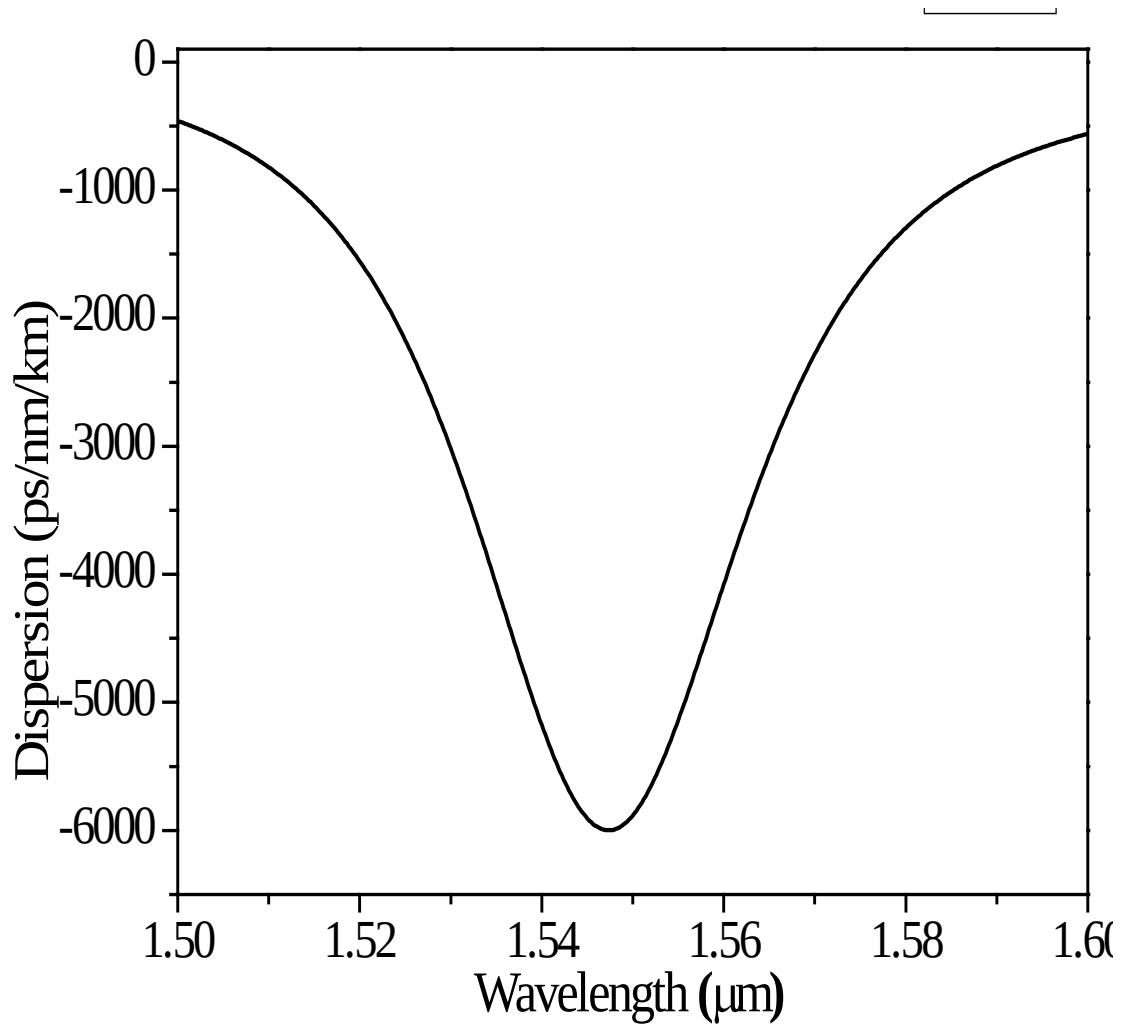

Fig. 5: Dispersion curves of PCF with Λ=1.20μm, $d_1$=0.96μm, and $d_2$=0.48μm.

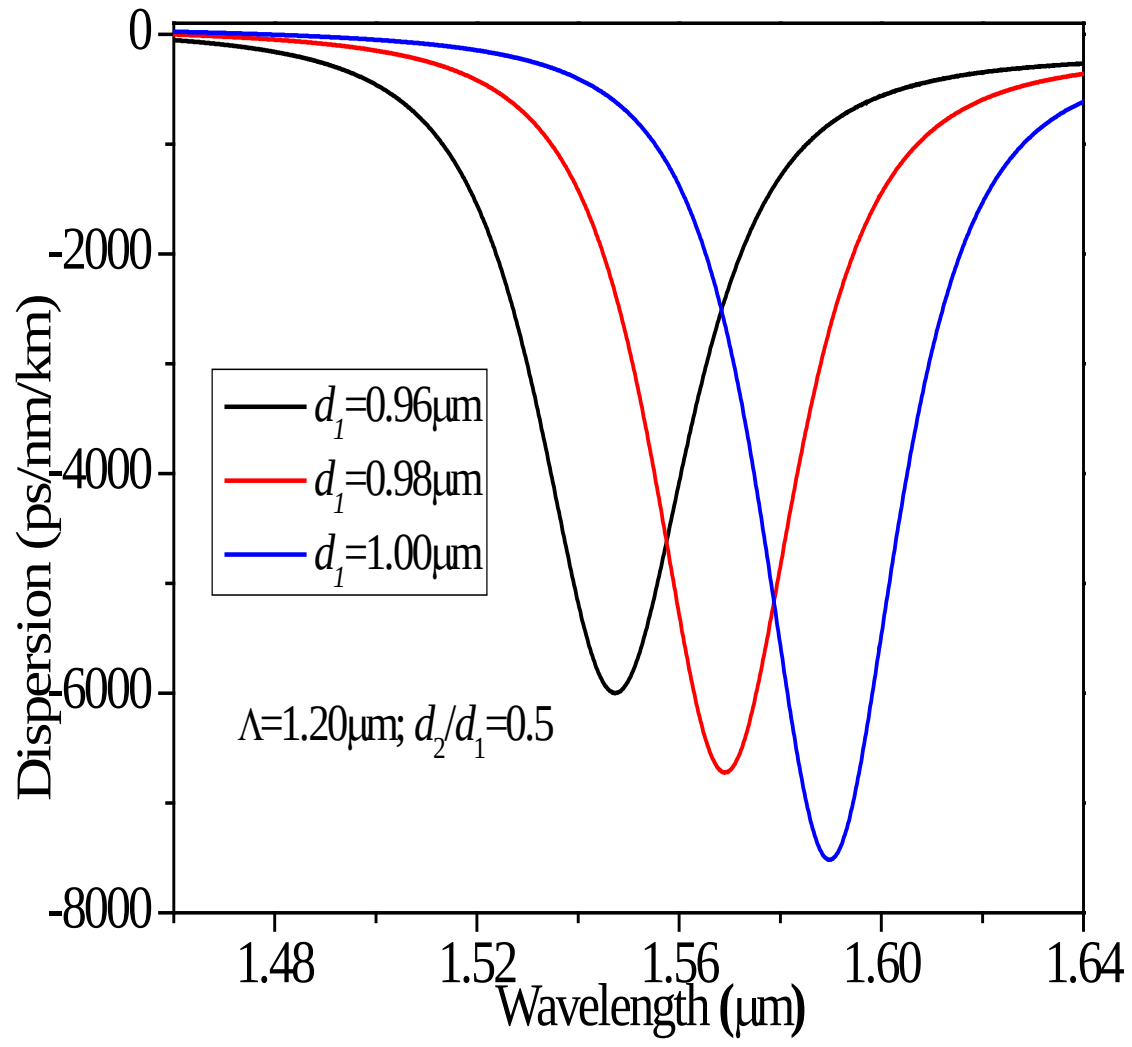

Fig. 6: Dispersion values for different values of $d_1$ keeping Λ=1.20μm and $d_2$/$d_1$=0.5

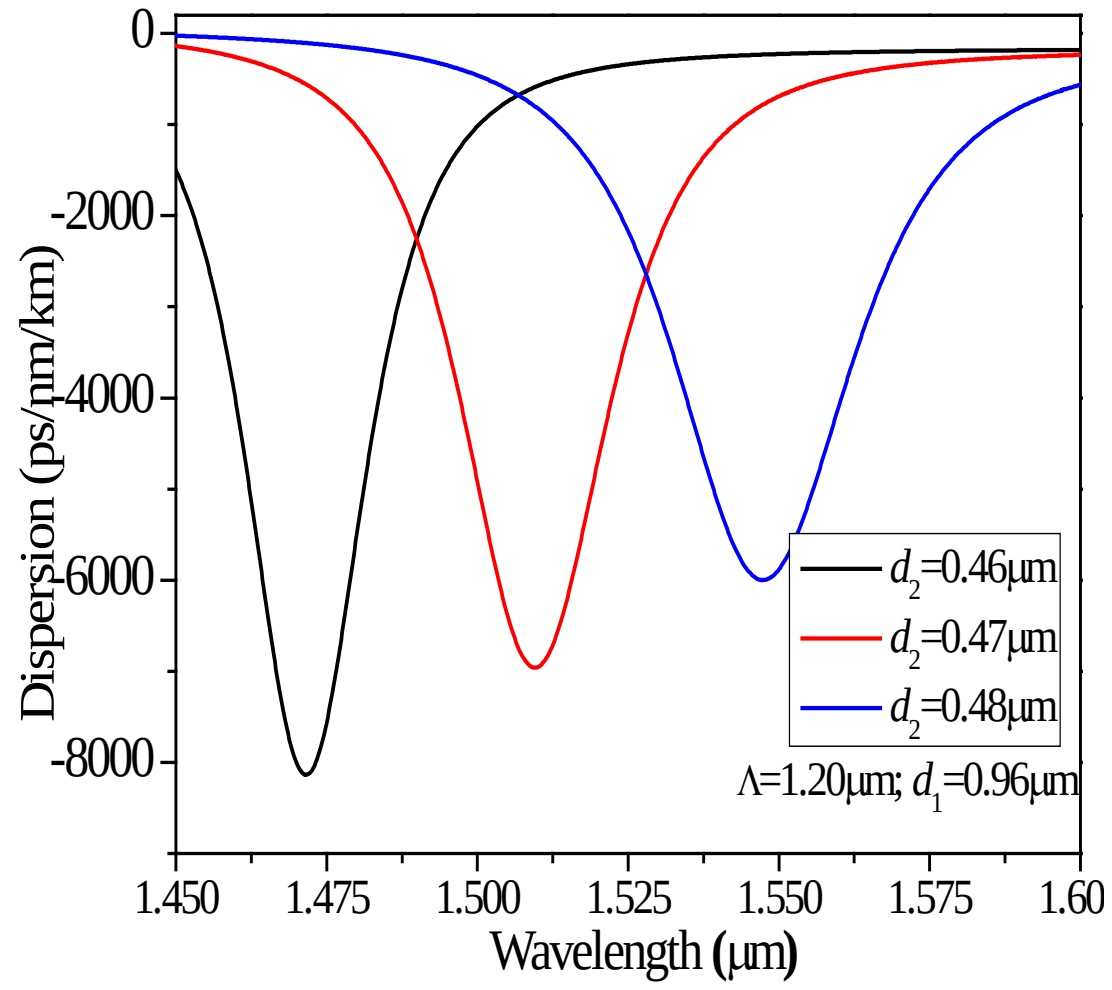

Fig. 7: Dispersion values for different values of $d_2$ keeping Λ=1.20μm and $d_1$=0.96μm

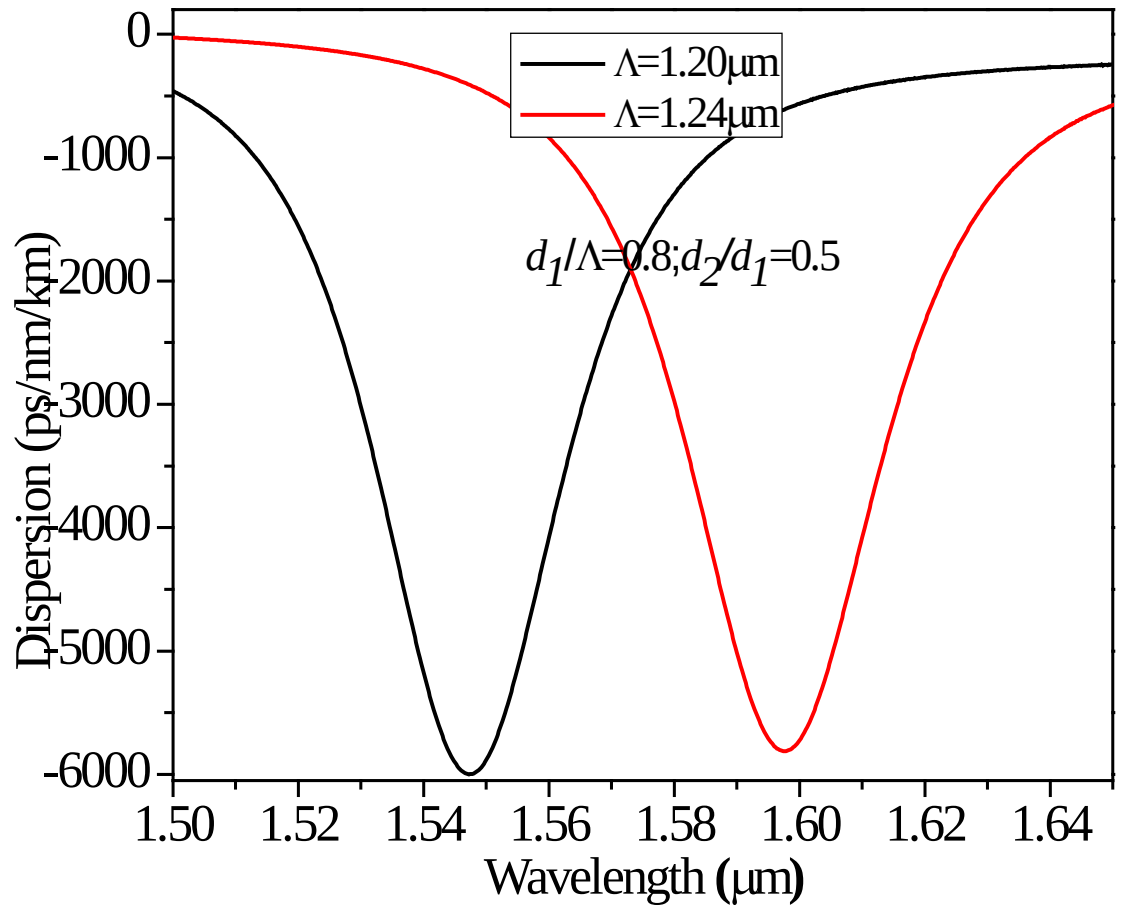

Fig. 8: Dispersion values for different values of Λ keeping $d_1$/Λ=0.8 and $d_2$/$d_1$=0.5

## 4: Design of Dispersion-Compensating Photonic Crystal Fiber

### 4.1: Large Negative Dispersion Narrowband Dispersion-Compensating Photonic Crystal Fiber

With the above discussion of the variation of the dispersion with the available structural parameters, we have achieved a dispersion values of −37,300 *ps/km/nm* around 1550 nm, with Λ=1.26 μm, $d_1$=1.17 μm and $d_2$=0.52 μm as shown in Fig. 9. While optimizing the design, we tried to keep the PCF parameters values such that the fabrication design is practical. The optimized dispersion value dispersion value can compensate for dispersion of 2000 times its length of standard single mode fiber used for long-distance optical data transmission systems. This value of negative dispersion with the regular PCF, which can be drawn with the regular stack and draw technique, is the highest to the best of our knowledge.

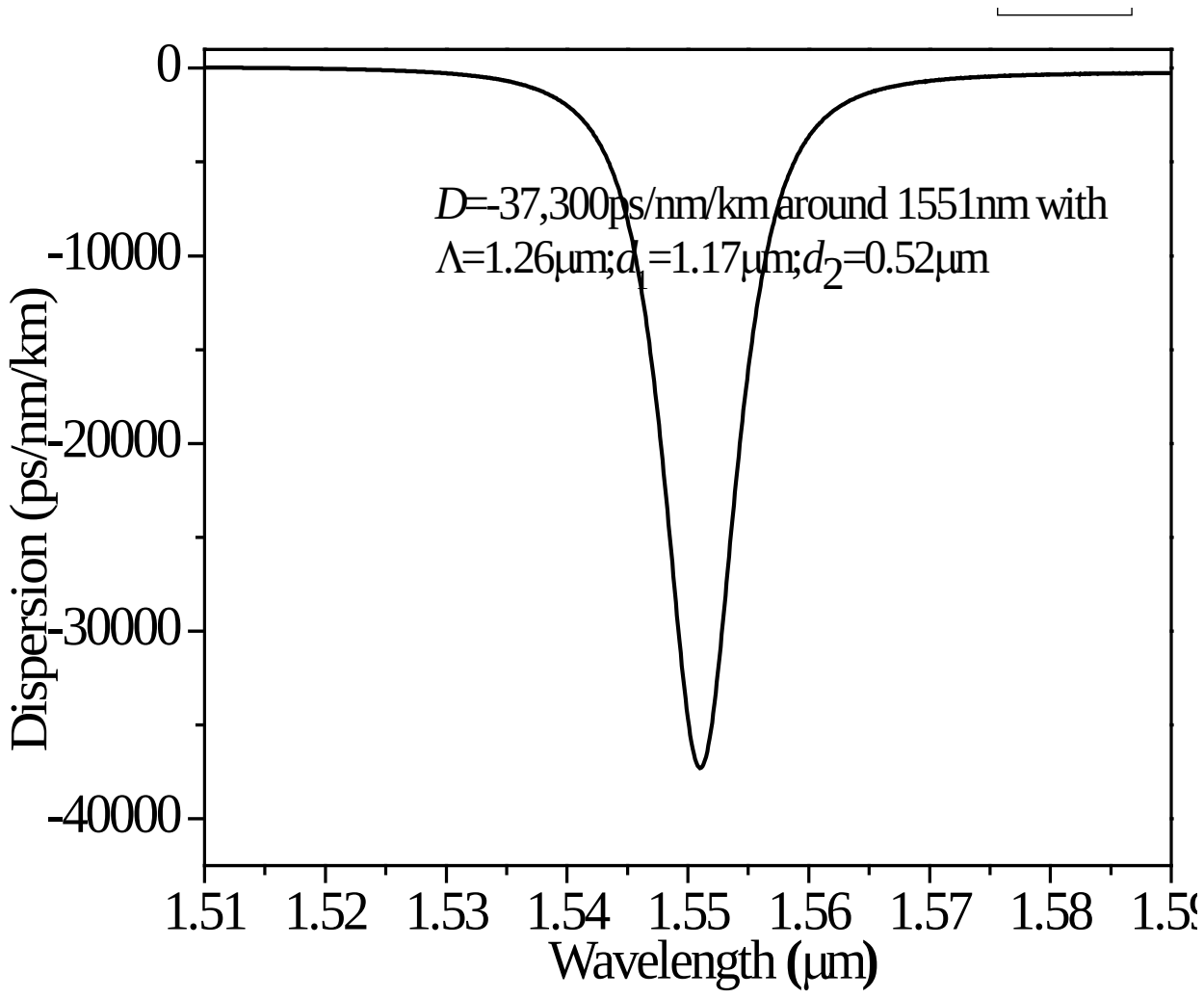

Fig. 9: High negative dispersion value of -37,300*ps/nm/km* has been achieved with Λ=1.24μm, $d_1$=1.17μm and $d_2$=0.52μm.

### 4.2: Broadband Dispersion-Compensating Photonic Crystal Fiber

For broadband dispersion compensation, we have considered $d_1$=0.76 µm and $d_2$=0.60 µm for different values of Λ. The dispersion values of PCF as a function of wavelength for the above parameters are shown in Fig. 10. In the spectral range of 1.4 – 1.6µm, dispersion values vary from −800 to −2000 *ps/km/nm* (Fig. 8(a)), the dispersion slope values are between −3$ps/km/nm^2$ and −6.5 $ps/km/nm^2$ (Fig. 8(b)) with kappa values varies between 200 nm to 300 nm (Fig. 8(c)), whereas it should be mentioned that the inline SMF-28 has kappa value of 280nm around the communication wavelength with dispersion values varying between 10 *ps/km/nm-20 ps/km/nm*. So the PCF can compensate for the dispersion of 100 times its length of standard single mode fiber used for broadband communication. Since the dispersion and the dispersion slope have similar kappa values, they can be compensated at the same time. The variation of dispersion, dispersion slope and kappa values for different $d_1$ with Λ=0.81 µm and $d_2$= 0.56 µm as a function of wavelength are shown in Fig. 11. The behavior of dispersion, dispersion slope and kappa as a function of wavelength of the graphs are similar to that of Fig. 10 with dispersion varies between −800 to −2500 *ps/km/nm* (Fig. 11(a)), the dispersion slope values are between −3.5$ps/km/nm^2$ and −9 $ps/km/nm^2$ (Fig. 11(b)) with kappa values varies around 250 nm (Fig. 11(c)). The variation of dispersion, dispersion slope and kappa values for different $d_2$ with Λ=0.82 µm and $d_1$= 0.76 µm values of PCF as a function of wavelength are shown in Fig. 12. The natures of the graphs are similar to that of Fig. 10 and Fig. 11 with dispersion varies between −800 to −2600 *ps/km/nm* (Fig. 12(a)), the dispersion slope values are between −3.5$ps/km/nm^2$ and −11 $ps/km/nm^2$ (Fig. 12(b)) with kappa values varies around 250 nm (Fig. 12(c)). With the above-mentioned methodology, best suitable inline dispersion compensating effect can be achieved by the associated PCF parameters. The present studies demonstrate the sensitivity of the parameters for fabrication tolerance. The investigations demonstrate that for broadband compensation, dispersion is insensitive to the change of structural parameters. Dispersion value changes almost linearly with wavelength for change of structural parameters ($d_1$, $d_2$ and Λ). So with the above fabrication tolerance analysis, broadband dispersion compensation PCF will be very easy to fabricate with the added fact that the PCF will be solely based upon pure silica and air-holes.

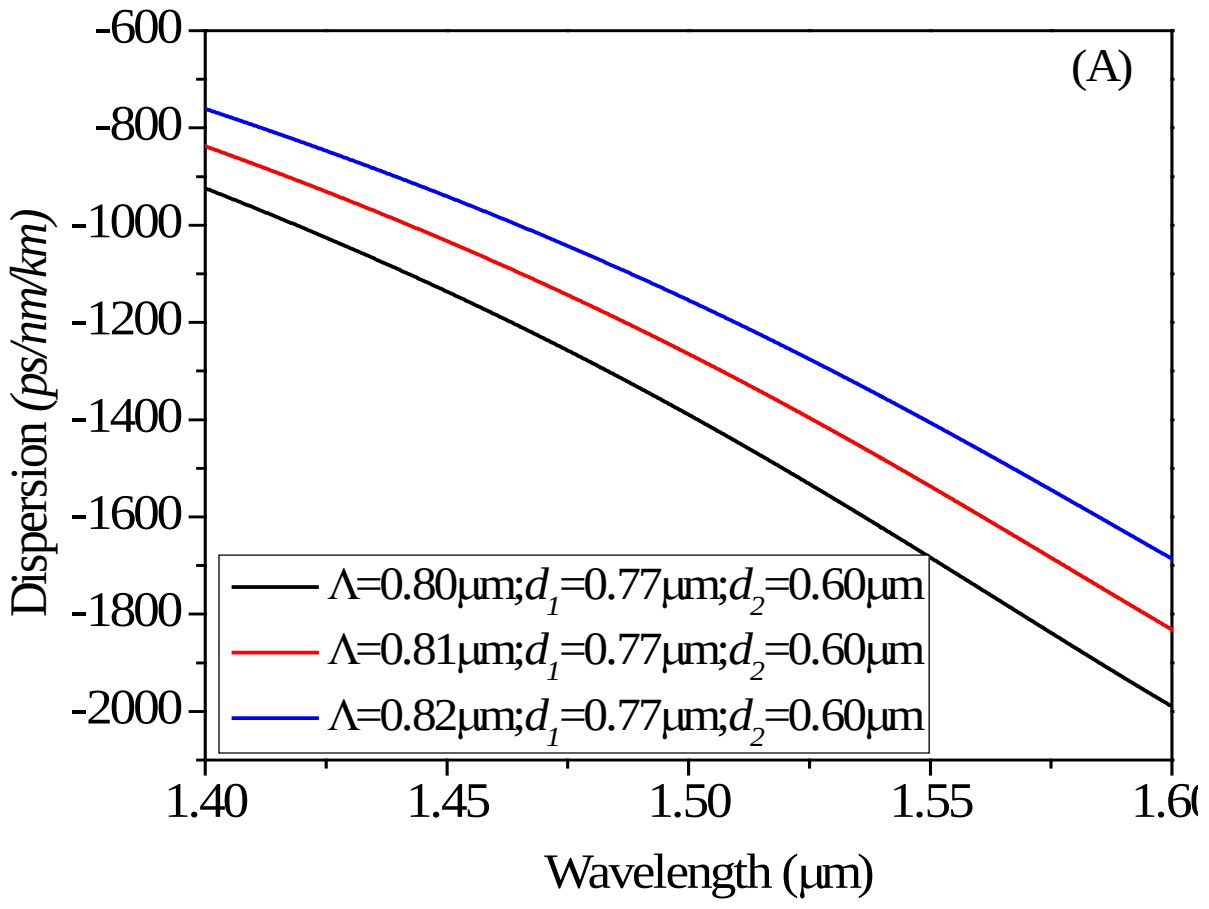

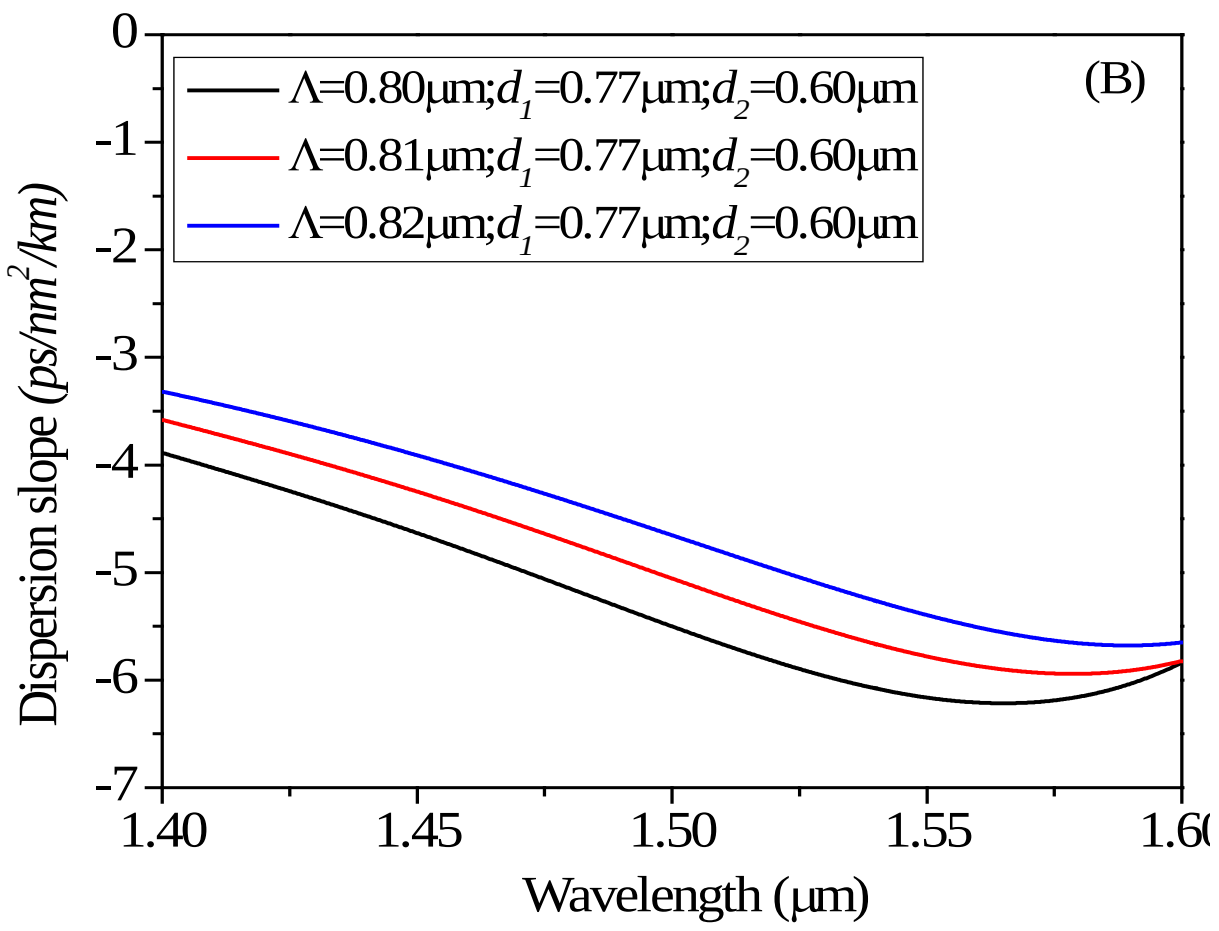

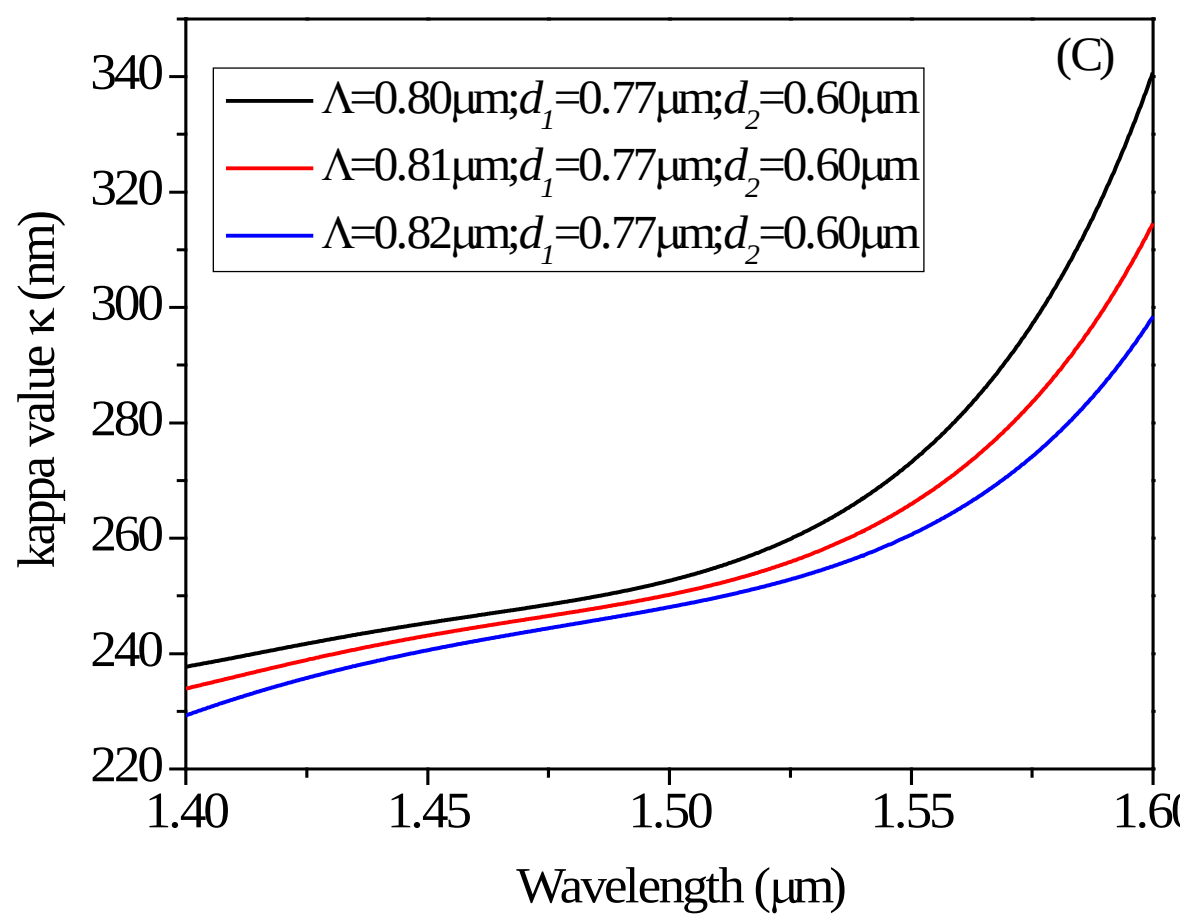

Fig. 10: (a) Dispersion (b) dispersion slope (c) kappa value of the PCF as a function of wavelength with for different values of Λ (0.80µm -- 0.82µm) for $d_1$ =0.77µm and $d_2$=0.60µm.

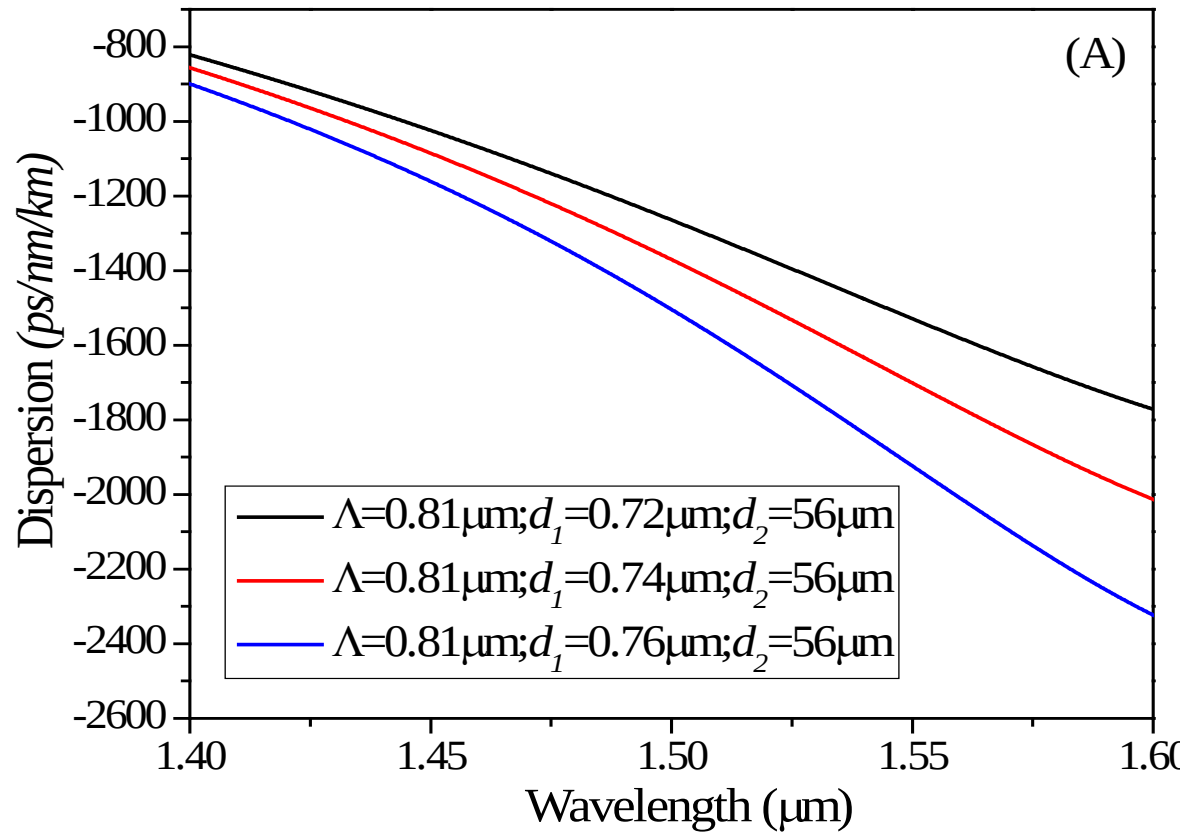

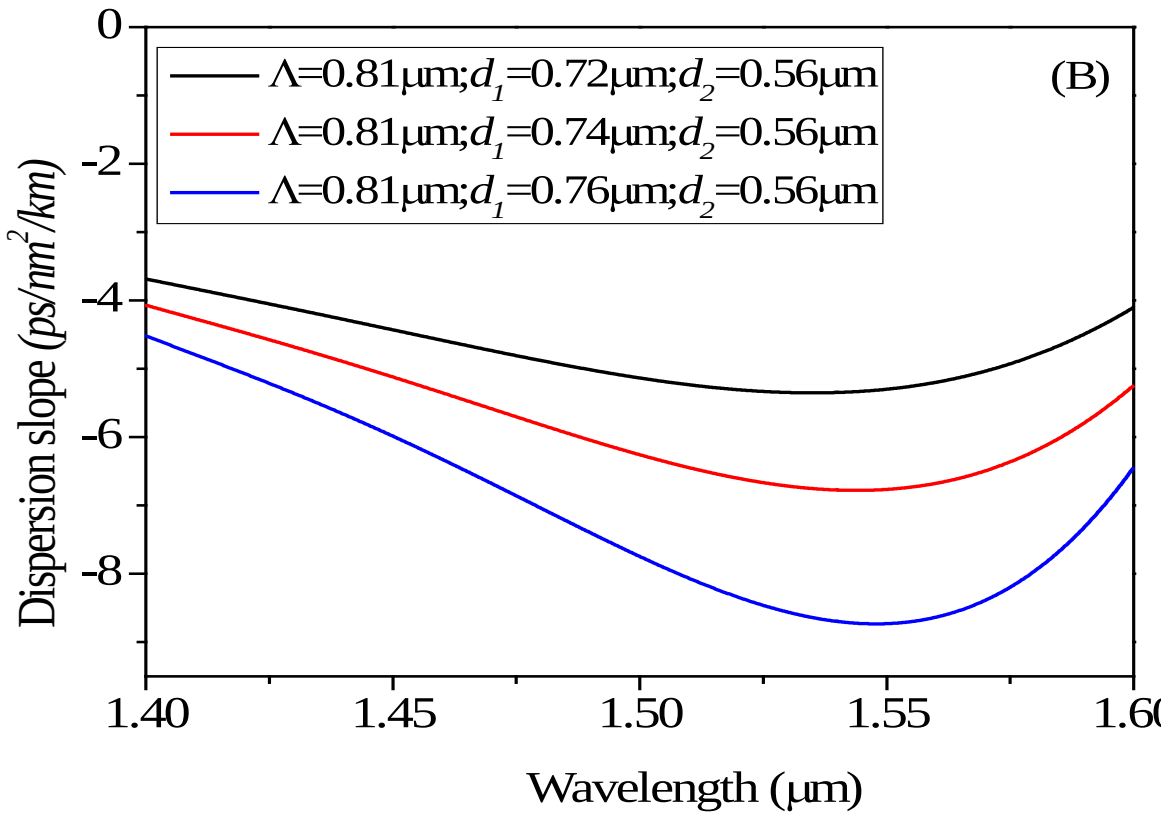

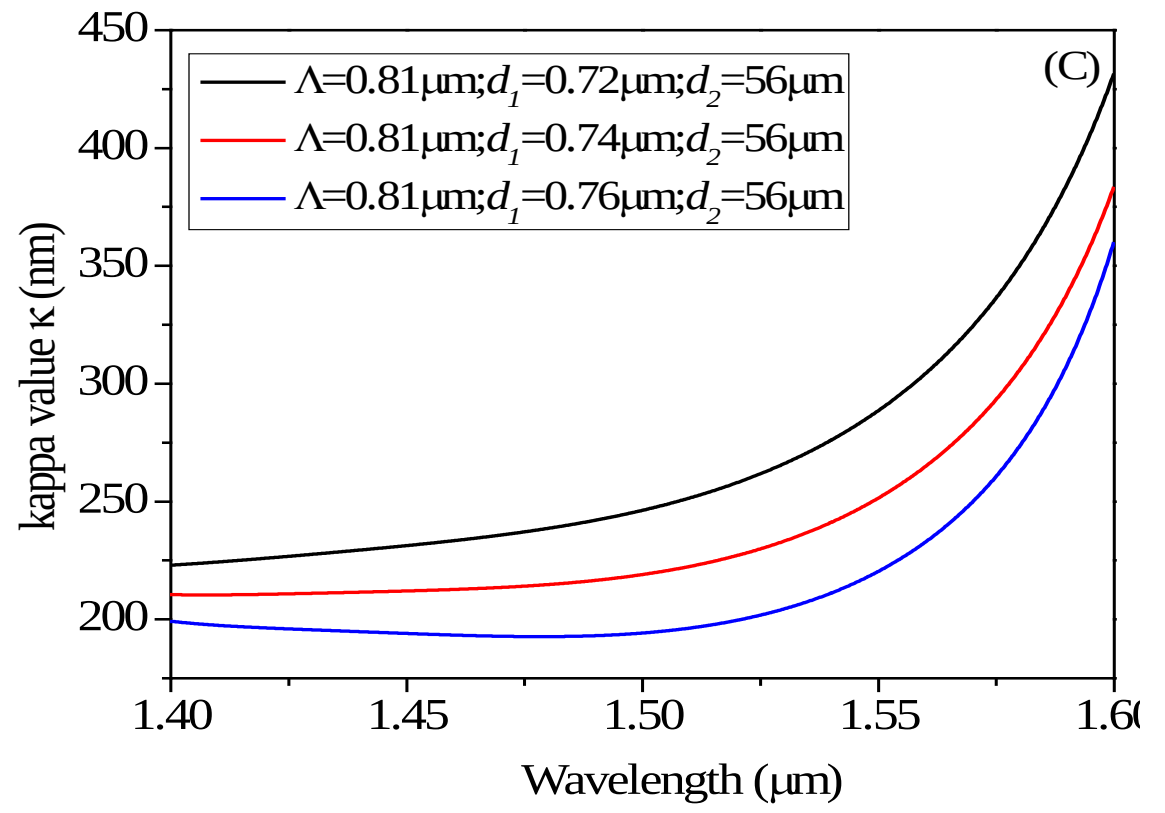

Fig. 11: (a) Dispersion (b) dispersion slope (c) kappa value of the PCF as a function of wavelength for different values of $d_1$ (0.72μm -0.76μm) for Λ=0.81μm and $d_2$=0.56μm.

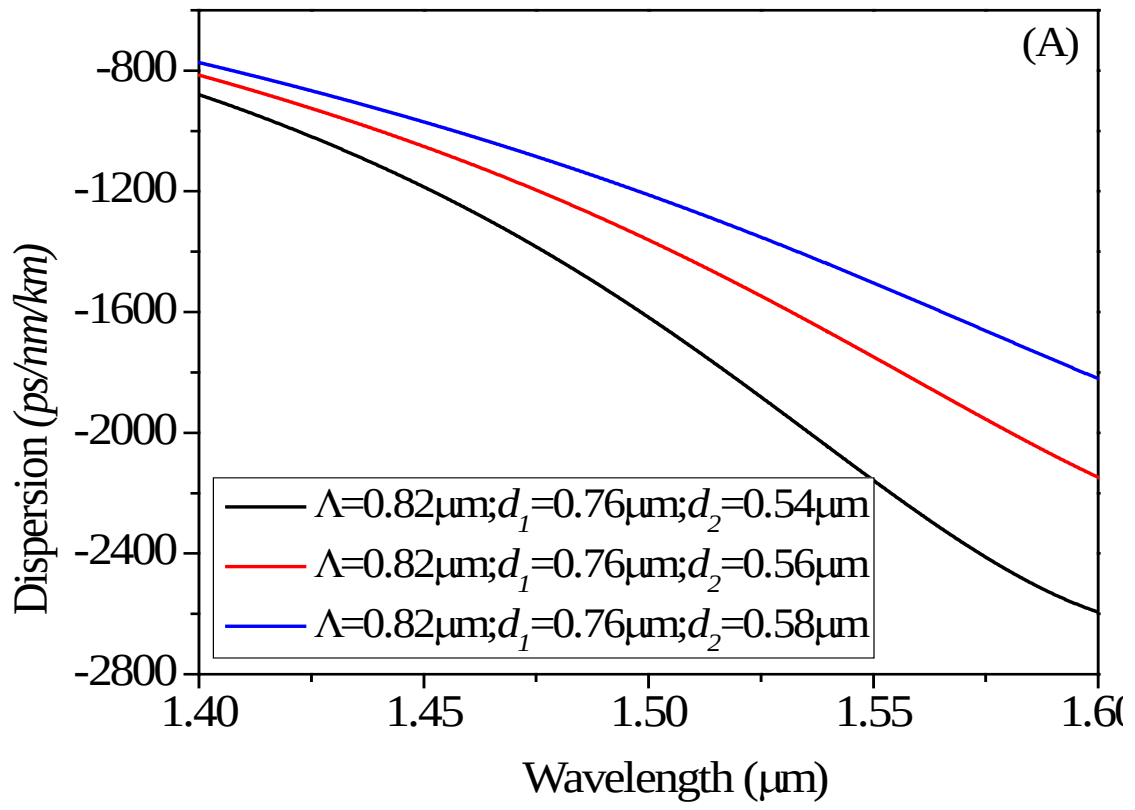

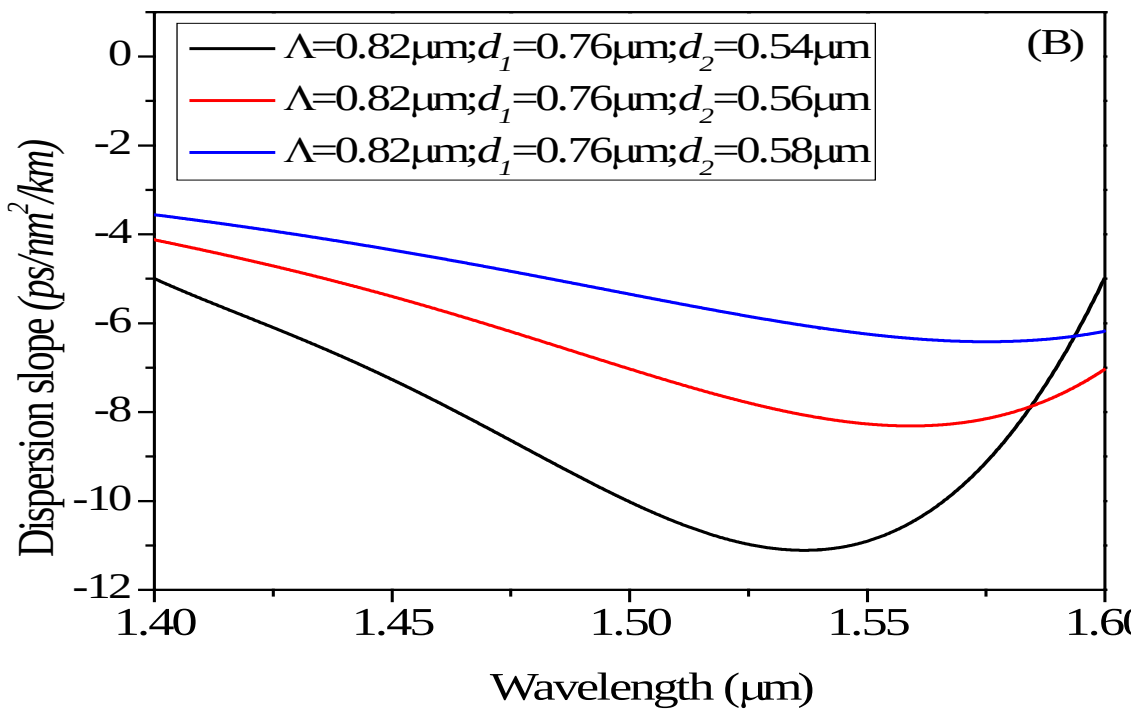

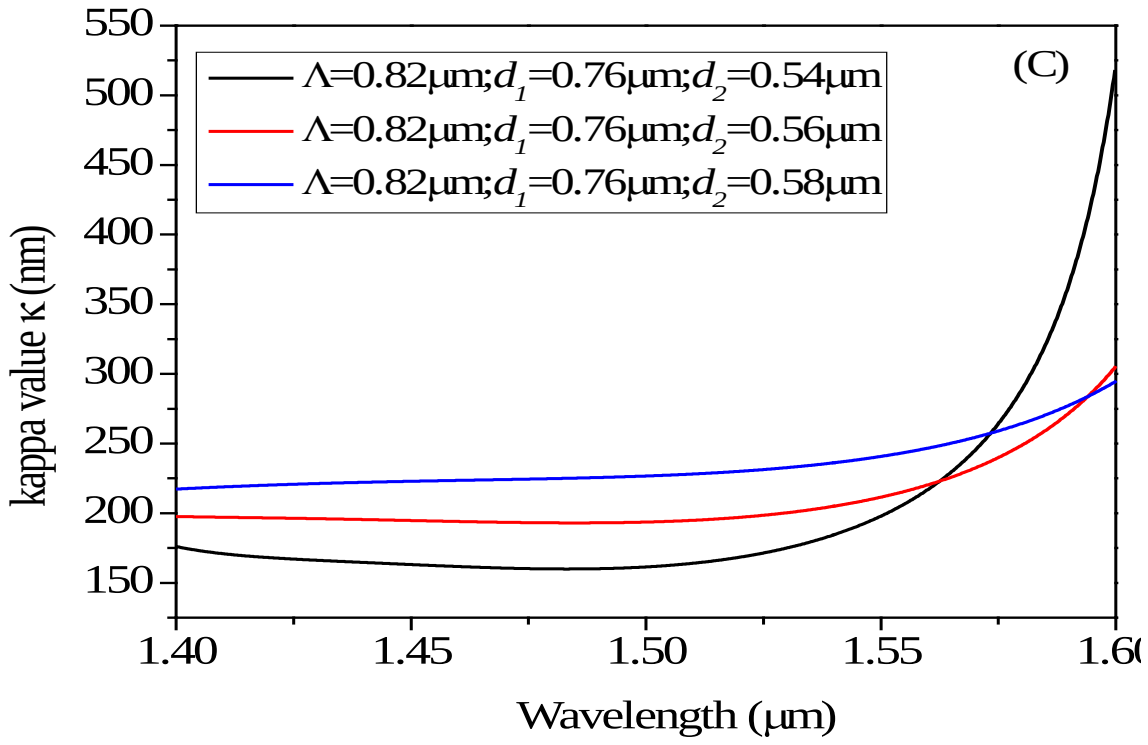

Fig. 12: (a) Dispersion (b) dispersion slope (c) kappa value of the PCF as a function of wavelength for different values of $d_2$ (0.54μm -0.58μm) for Λ=0.82μm and $d_1$=0.76μm.

**5. ASE suppression with the dual core PCF:**

Amplified Spontaneous Emission (ASE) is responsible for reasonable gain in a doped fiber amplifier. Due to strong confinement of the mode in the core region, a strong overlap between the optical field and the doped region (generally in the core) can be anticipated which consequently amplifies the input signal. However, after a certain wavelength ASE can be drastically reduced [21-26] and therefore it can be applicable for different applications like S band (1460 nm to 1530 nm) amplifier with ASE, after the required wavelength, drastically reduced. Thus, one can engineer the PCF geometrical parameters to design required performance in terms of gain and ASE suppression.

We have calculated the overlap factor given by Eqn. 4 which is defined as the overlap between the doped amplifying ions (here the Erbium ions) and signal optical mode field and is calculated as reported in [27, 28]

$$\Gamma_x = \frac{\int_0^{r_d}\int_0^{2\pi}\left|E_x^2(x,y)\right|dxdy}{\int_0^{\infty}\int_0^{2\pi}\left|E_x^2(x,y)\right|dxdy} \quad (4)$$

where $x$=s(signal) or $p$ (pump), $r_d$ is the radius of erbium-doped area and $\left|E_x^2(x,y)\right|$ is the electric field intensity of the guided wave at $\lambda_x$ ($x$=s or p). We numerically compute the mode-field distribution of the optimized concentric dual-core PCF structure as shown in Fig. 9. In this calculation, we have considered that the doped radius to be equal to the Λ/2 of the PCF. Figure 13 gives the overlap factor for the optimized fiber presented in Fig. 9 with Λ=1.26 μm, $d_1$=1.17 μm and $d_2$=0.52 μm with the coupling wavelength of 1551 nm.

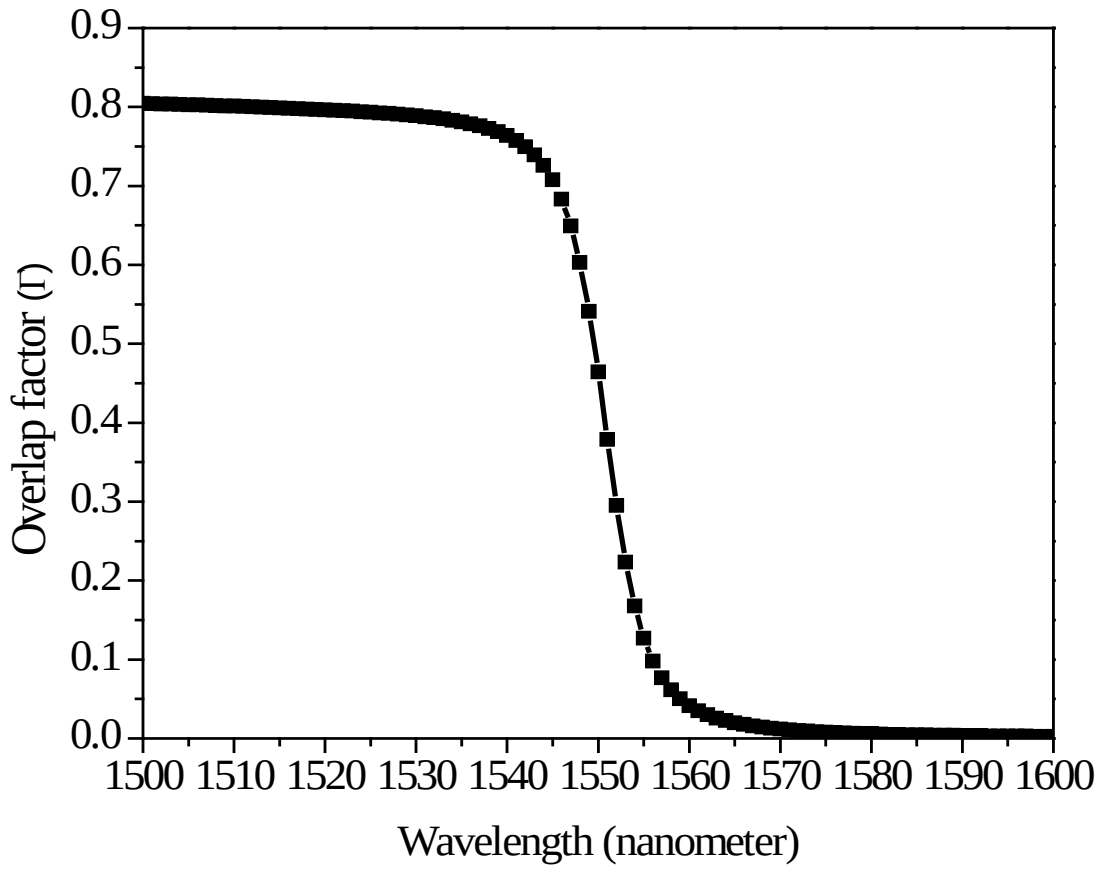

Fig. 13: Overlap factor variation of the optimized PCF shown in Fig. 9.

Looking at the overlap factor variation for the above optimized fiber, it can be observed that the overlap factor is 80% around λ=1.54 μm and then starts decreasing as the wavelength approaches the coupling wavelength, then sharply drops below 4% at around λ =1.56 μm, and becomes almost zero for longer wavelengths. This suggests that more than 96% of the optical power of the inner core mode is transferred to the outer ring core. As the active erbium ions remain confined in the central core, it is expected that all the wavelengths will be attenuated after 1.56 μm. Therefore, we can get high overlap factor values for all wavelengths below the coupling wavelength and less than 4% of the overlap factor for all wavelengths after 1.56 μm for the optimized PCF structure.

By proper design of the available parameters and with the above methodology we have presented, we can change the coupling wavelength to any values according to the requirement. If the coupling wavelength is fixed around 1530 nm we can have one S band ASE suppressed amplifier which will amplify only the wavelength below 1530 nm.

## 6: Conclusions

In the present work, we report that for efficient dispersion compensation the effect of DCFs can be realized in a PCF with two different types of air-hole diameters in the cladding. The dependence of different PCF parameters upon the dispersion has been studied in details. Based upon the findings of the individual parameter dependence of the PCFs upon dispersion, both narrowband and broadband dispersion compensation can been achieved. Our numerical analysis establishes that, by proper adjustment of the structural parameters, one can obtain a narrowband DCF with dispersion values of −37,300 *ps/km/nm*, around 1.55 μm of wavelength. This optimized value of negative dispersion is the maximum to the best of our knowledge that can be drawn with regular stack and drawn technique. The optimized fiber can compensate the dispersion of thousand times the length of the standard single mode fiber. The dual core PCF was also found to be suitable for broadband dispersion compensation with dispersion values from −1000 to −2600 *ps/km/nm* over the 200 nm range and kappa values near 300 nm, which matched well with the inlaying standard single mode fiber. The designed fiber can compensate for the inline dispersion few hundred times its length of standard single mode fibers and can be used for dispersion compensation in long-haul data transmission and broadband dispersion compensation. Finally, we have applied the principle of power transfer from inner core to the outer core of dual core based structure around the coupling wavelength, to design a possible ASE suppressed amplifier, where wavelengths after the coupling wavelength cannot be amplifier as most of the power tunnels to the outer core after the coupling wavelength.

**Acknowledgement:**

The authors would like to thank Dr. Boris Kuhlmey, University of Sydney, Australia for providing valuable suggestions in understanding the software for designing and studying the properties of different structures. The authors acknowledge sincerely the Defence Research and Development Organization, Govt. of India and CRF of IIT Kharagpur for the financial support to carry out this research.

**References:**

[1]. K. Thyagarajan, R. K. Varshney, P. Palai, A. K. Ghatak, and I. C. Goyal, “A novel design of a dispersion compensating fiber,” IEEE Photon. Technol. Lett. 8, 1510–1512 (1996).

[2]. J.-L. Auguste, R. Jindal, J.-M. Blondy, M. Clapeau, J. Marcou, B. Dussardier, G. Monnom, D. B. Ostrowsky, B. P. Pal, and K. Thyagarajan, “−1800 ps/(nm km) chromatic dispersion at 1.55 μm in dual concentric core fibre,” Electron. Lett. 36, 1689–1691 (2000).

[3]. J. L. Auguste, J. M. Blondy, J. Maury, J. Marcou, B. Dussardier, G. Monnom, R. Jindal, K. Thyagarajan, and B. P. Pal, “Conception, realization, and characterization of a very high negative chromatic dispersion fiber,” Opt. Fiber Technol. 8, 89–105 (2002).

[4]. L. Grüner-Nielsen, M. Wandel, P. Kristensen, C. Jorgensen, L. Jorgensen, B. Edvold, B. Pálsdóttir, and D. Jakobsen, “Dispersion-compensating fibers,” J. Lightwave Technol. 23, 3566–3579 (2005).

[5]. F. Gérôme, J. Auguste, J. Maury, J. Blondy, and J. Marcou, “Theoretical and experimental analysis of a chromatic dispersion compensating module using a dual concentric core fiber,” J. Lightwave Technol. 24, 442–448 (2006).

[6]. G. Ouyang, Y. Xu, and A. Yariv, “Theoretical study on dispersion compensation in air-core Bragg fibers,” Opt. Express 10, 899–908 (2002).

[7]. T. D. Engeness, M. Ibanescu, S. G. Johnson, O. Weisberg, M. Skorobogatiy, S. Jacobs, and Y. Fink, “Dispersion tailoring and compensation by model interactions in OmniGuide fibers,” Opt. Express 11, 1175–1196 (2003).

[8]. C. D. Poole, J. M. Wiesenfeld, D. J. DiGiovanni, and A. M. Vengsarkar, “Optical fiber-based dispersion compensation using higher order modes near cutoff,” J. Lightwave Technol. 12, 1746–1758 (1994).

[9]. S. Ramachandran, B. Mikkelsen, L. C. Cowsar, M. F. Yan, G. Raybon, L. Boivin, M. Fishteyn, W. A. Reed, P. Wisk, D. Brownlow, R. G. Huff, and L. Gruner-Nielsen, “All-fiber grating-based higher order mode dispersion compensator for broad-band compensation and 1000 km transmission at 40Gb/s,” IEEE Photon. Technol. Lett. 13, 632–634 (2001).

[10]. S. Ghalmi, S. Ramachandran, E. Monberg, Z. Wang, M. Yan, F. Dimarello, W. Reed, P. Wisk, and J. Fleming, “Low-loss, all-fibre higher-order-mode dispersion compensators for lumped or multi-span compensation,” Electron. Lett. 38, 1507–1508 (2002).

[11]. J. Broeng, D. Mogilevstev, S. E. Barkou and A. Bjakle, “Photonic

Crystal Fibers: a new class of optical waveguides" Opt. Fiber Tech. **5**, 305-330 (1999).

[12]. P. St. J. Russel, "Photonic-Crystal Fibers". J of Lightwave Tech. **24**, 4729-4749 (2006).

[13]. T. A. Birks, D. Mogilevtsev, J. C. Knight, P. St, and J. Russel, "Dispersion compensation using single material fibers," IEEE Photon. Technol. Lett. 11, 674–676 (1999).

[14]. L. P. Shen, W.-P. Huang, and S. S. Jian, "Design of photonic crystal fibers for dispersion-related applications," J. Lightwave Technol. 21, 1644–1651 (2003).

[15]. Y. Ni, L. An, J. Peng, and C. Fan, "Dual-core photonic crystal fiber for dispersion compensation," IEEE Photon. Technol. Lett. 16, 1516–1518 (2004).

[16]. A. Huttunen and P. Törmä, "Optimization of dual-core and microstructure fiber geometries for dispersion compensation and large mode area," Opt. Express 13, 627–635 (2005).

[17]. S. Yang, Y. Zhang, L. He, and S. Xie, "Broadband dispersion-compensating photonic crystal fiber," Opt. Lett. 31, 2830–2832 (2006).

[18]. CUDOS MOF utilities available online: http://www.physics.usyd.edu.au/cudos/mofsoftware/

[19]. T. P. White, B. T. Kuhlmey, R. C. PcPhedran, D. Maystre, G. Renversez, C. M de Sterke and L. C. Botten, "Multipole method for microstructured optical fibers. I. Formulation*" J. Opt. Soc. Am. B.* **19**, 2322 (2002).

[20]. B. T. Kuhlmey, T. P. White, R. C. PcPhedran, D. Maystre, G. Renversez, C. M de Sterke and L. C. Botten, "Multipole method for microstructured optical fibers. II. Implementataion and results." *J. Opt. Soc. Am. B.* **19**, 2331 (2002).

[21]. H. Ono, M. Yamada, and M. Shimizu, "S-band erbium doped fiber amplifiers with a multistage configuration-design, characterization, and gain tilt compensation," *J. Lightw. Technol.*, vol. 21, no. 10, pp. 2240–2246, Oct. 2003.

[22]. K. Thyagarajan and C. Kakkar, "S-band single-stage EDFA with 25-dB gain using distributed ASE suppression," *IEEE Photon. Technol. Lett.*, vol. 16, no. 11, pp. 2448–2450, Nov. 2004.

[23]. J. B. Rosolem, A. A. Juriollo, R. Arradi, A. D. Coral, J. C. R. F. Oliveira, and M. A. Romero, "All silica S-band double-pass configuration," *IEEE Photon. Technol. Lett.*, vol. 17, no. 7, pp. 1399–1401, Jul. 2005.

[24]. M. Foroni, F. Poli, A. Cucinotta, and S. Selleri, "S-band depressed-cladding erbium-doped fiber amplifier with double-pass configuration," *Opt. Lett.*, vol. 31, no. 22, pp. 3228–3230, Nov. 2006.

[25]. J. B. Rosolem, A. A. Juriollo, and M. A. Romero, "S-band EDFA using standard erbium-doped fibre," *Electron. Lett.*, vol. 43, no. 22, pp. 1186–1188, Oct. 2007.

[26]. L. Vincetti, M. Foroni, F. Poli, M. Maini, A. Cucinotta, S. Selleri, and M. Zoboli, "Numerical modeling of S-band EDFA based on distributed fiber loss," *J. Lightw. Technol.*, vol. 26, no. 14, pp. 2168–2174, Jul. 2008.

[27]. S. Hilaire, D. Pagnoux, P. Roy, and S. Fevrier, "Numerical study of single-mode Er-doped microstructured fibers: Influence of geometrical parameters on amplifier performances," *Opt. Exp.*, vol. 14, no. 22, pp. 10865–10877, Jul. 2006.

[28]. S. K. Varshney, , K. Saitoh, , K., Masanori;.; R. K. Sinha,.; B. P. Pal; "Design of S-Band Erbium-Doped Concentric Dual-Core Photonic Crystal Fiber Amplifiers With ASE Suppression", *Journal of Lightwave Technology*, Volume.27, Issue.11, pp.1725, 2009, ISSN: 07338724,